\documentclass[10pt,a4paper,oneside,english]{elsart}
\usepackage[T1]{fontenc}
\usepackage[latin1]{inputenc}
\usepackage{float}
\usepackage{graphicx}
\usepackage{amssymb}

\usepackage{babel}
\usepackage{epsfig}
\makeatother
\begin{document}
\begin{frontmatter}

\title{The Evolution of Interdependence in World Equity Markets - Evidence
from Minimum Spanning Trees}

\author[TCD_Physics]{Ricardo Coelho},
\ead{rjcoelho@tcd.ie}
\author[Kings]{Claire G. Gilmore},
\author[TCD_Bus]{Brian Lucey},
\author[TCD_Physics]{Peter Richmond}
\author[TCD_Physics]{\and Stefan Hutzler}

\address[TCD_Physics]{School of Physics, Trinity College Dublin, Dublin 2, Ireland}
\address[Kings]{McGowan School of Business, King's College, Wilkes-Barre, PA, U.S.A.}
\address[TCD_Bus]{Institute for International Integration Studies and School of Business,
Trinity College Dublin, Dublin 2, Ireland}

\begin{abstract}
The minimum spanning tree is used to study the process of market integration for a large group of national stock market indices. We show how the asset tree evolves over time and describe the dynamics of its normalized length, mean occupation layer, and single- and multiple-step linkage survival rates. Over the period studied, 1997-2006, the tree shows a tendency to become more compact. This implies that global equity markets are increasingly interrelated. The consequence for global investors is a potential reduction of the benefits of international portfolio diversification.
\end{abstract}

\end{frontmatter}

\section{Introduction}

This paper examines the extent and evolution of interdependence between world equity markets over a $10$-year period using the Minimum Spanning Tree (MST) approach of Mantegna \cite{Mantegna_EPJB11_193_1999}. The approach derives ultimately from graph theory and has been used as a simple way to study the correlations of stocks in a stock market. One advantage that MST analysis has over traditional finance perspectives on international equity market integration is that it provides a parsimonious representation of the network of all possible interconnectedness. With $N$ equity indices the number of possible nodal connections is large, $N(N-1)/2$.  The MST can greatly reduce this complexity and show only the $N - 1$ most important non-redundant connections in a graphical manner. The MST approach also provides useful information in terms of the centrality or otherwise of individual equity markets (nodes) in the overall system.

A large body of research exists in the finance literature on the integration of international equity markets. There are three basic approaches to defining the extent to which international financial markets are integrated. These fall into two broad categories, direct measures and indirect measures. The first approach, a direct measure, is couched in terms of the extent to which the rates of return on financial assets with similar risk characteristics and maturity are equalized across political jurisdictions. We call this a direct measure as it invokes the so-called law of one price. The second approach involves the concept of international capital market completeness and in essence examines the extent to which asset prices and returns are a function of domestic or international factors. The third approach is based on the extent to which domestic investment is financed from world savings rather than from domestic savings. Both of these latter measures can be called indirect.

In general, where possible, finance researchers consider the direct approach to be preferable \cite{Kearney_IRFA13_571_2004}. The direct approach to integration has been studied by a number of researchers who have evaluated the evolution of equity market correlations, the extent to which common stochastic trends in returns emerge, and the specification of dynamic paths toward greater integration between the returns on equities.  Surveys of this literature can be found in \cite{Kearney_IRFA13_571_2004,Bekaert,Goetzman}. We contribute to the research on the dynamic process of integration through the use of MST analysis, which is particularly suitable for extracting the most important information when a large number of markets are under examination.

MST analysis has been applied previously to analyze the clustering behavior of individual stocks within a single country, usually the US \cite{Bonanno_QF1_96_2001,Bonanno_PRE68_046130_2003,Bonanno_EPJB38_363_2004,Vandewalle_QF1_372_2001}. These studies typically find a strong correspondence between business sector and cluster structure, illustrating the ability of the MST methodology to convey meaningful economic information. While these are static analysis, a variety of dynamic analysis of the time-varying behavior of stocks has also been developed in \cite{Onnela_EPJB30_285_2002,Onnela_PhysA324_247_2003,Onnela_PST48_106_2003,Onnela_PRE68_056110_2003,Micciche_PhysA324_66_2003,Coelho_PH0601189}. MST analysis has also been applied to the foreign exchange markets as a means to trace the dynamics of relationships between currencies \cite{McDonald_PRE72_046106_2005}.

To our knowledge only one study has been published to date applying the MST approach to groups of national equity markets \cite{Bonanno_PRE62_7615_2000}. There is one other earlier study \cite{Panton_JFQA11_415_1976}, which however used clustering analysis. A simple dynamic analysis based on partially overlapping windows of indices for $20$ countries for the years 1988-1996 finds that markets group according to a geographical principle, as is also the case for a static examination of $51$ world indices for the years 1996-1999 in the same study \cite{Bonanno_PRE62_7615_2000}. Our research significantly extends this work by applying dynamic MST methods to examine  the time-varying behavior of global equity market co-movements for a group of $53$  developed, emerging and developing countries over the years 1997-2006. This period includes major market events such as the Asian and Russian economic crises, the introduction of the euro, and the enlargement of the European Union (EU). In addition to confirming the earlier evidence of a geographical organizing principle we document a tendency of the MST toward higher density over time, indicating an increasing degree of integration of international equity markets. Such a finding is of interest to portfolio managers and investors, as the implication is of decreased potential for diversification benefits and thus perhaps decreased returns for international investors.

\section{Data}

We analyze the returns on $53$ countries' equity markets. The data consist of Morgan Stanley Capital International (MSCI) daily closing price indices for $44$ countries, for the period January 8, 1997, through February 1, 2006. An additional nine countries are also included in the sample, for a total of $53$. These countries and indices are: Croatia (Nomura), the Czech Republic (PX 50), Hungary (BUX), Iceland (ICEX 15 Cap), Lithuania (Nomura), Malta (HSBC Bank), Romania (Nomura), Slovakia (SAX) and Slovenia (HSBC Bank). All series are expressed in US dollar terms as the reference currency, thus reflecting the perspective of an international investor. All data are sourced from DataStream, Thomson Financial. One issue that needs to be addressed is the non-synchronous nature of the data, that is  the fact that equity markets open at different times. Recent research suggests that the use of daily data may lead to significant underestimation of equity market integration \cite{Schotman_WP04_017_2004}. As a consequence, to minimize the problem of non-synchronous trading the daily index level data were converted to weekly (Wednesday) returns:  $R_{i,t}=\ln(P_{i,t}/P_{i,t-1})$, where $P_{i,t}$ is the closing price of index $i$ at time $t$. The resulting number of weekly observations is $475$. 
The $53$ countries in our study and the respective symbols are represented in Table \ref{Table1}.

\begin{table}[h!]
\caption{Coutries and respective symbol.}
\centering
\begin{minipage}{\textwidth}
\centering
\begin{small}\sffamily
\begin{tabular}{|c|c||c|c||c|c|}
\hline 
Symbol&
Country&
Symbol&
Country&
Symbol&
Country\tabularnewline
\hline
\hline 
ARG&
Argentina&
HUN&
Hungary&
PHI&
Philippines\tabularnewline
\hline 
AUS&
Australia&
ICE&
Iceland&
POL&
Poland\tabularnewline
\hline 
AUT&
Austria&
IDO&
Indonesia&
PRT&
Portugal\tabularnewline
\hline 
BEL&
Belgium&
IND&
India&
ROM&
Romania\tabularnewline
\hline 
BRZ&
Brazil&
IRE&
Ireland&
RUS&
Russia\tabularnewline
\hline 
CAN&
Canada&
ISR&
Israel&
SAF&
South Africa\tabularnewline
\hline 
CHF&
Switzerland&
ITA&
Italy&
SGP&
Singapore\tabularnewline
\hline 
CHL&
Chile&
JAP&
Japan&
SOK&
South Korea\tabularnewline
\hline 
COL&
Colombia&
JOR&
Jordan&
SVK&
Slovakia\tabularnewline
\hline 
CRT&
Croatia&
LTU&
Lithuania&
SVN&
Slovenia\tabularnewline
\hline 
CZK&
Czech Republic&
MAL&
Malaysia&
SWE&
Sweden\tabularnewline
\hline 
DNK&
Denmark&
MEX&
Mexico&
THI&
Thailand\tabularnewline
\hline 
ESP&
Spain&
MTA&
Malta&
TUK&
Turkey\tabularnewline
\hline 
FIN&
Finland&
NEZ&
New Zealand&
TWA&
Taiwan\tabularnewline
\hline 
FRA&
France&
NLD&
Netherlands&
UK&
United Kingdom\tabularnewline
\hline 
GER&
Germany&
NOR&
Norway&
USA&
United States\tabularnewline
\hline 
GRC&
Greece&
PAK&
Pakistan&
VEZ&
Venezuela\tabularnewline
\hline 
HK&
Hong Kong&
PER&
Peru&
\multicolumn{2}{c}{}\tabularnewline
\cline{1-1} \cline{2-2} \cline{3-3} \cline{4-4} 
\end{tabular}
\end{small}
\end{minipage}
\label{Table1}
\end{table}

The reliance for the most part on MSCI indices allows for significant confidence in the findings, as these indices are designed explicitly to allow for cross market consideration of returns by investors. By contrast, studies that rely on indices from the individual equity markets, indices such as the NIKKEI225, the DJIA or the FTSE100, run the risk of non-comparability due to differences in construction, coverage and completeness.

\section{Methodology}

The MST is a connected graph in which each (random) variable in a data set, here a set of equity indices, is represented by a node. All the $N$ nodes are connected with $N-1$ links such that no loops are created. The correlation matrix of the data set, converted to an appropriate metric, measures the distances between indices, or nodes. The nodes are connected in the MST such that the sum of all distances is minimized.

Construction of the MST begins with a computation of the correlations between all pairs of weekly returns in the dataset of $53$ equity indices. The correlation coefficient is calculated between all pairs $(i,j)$ as follows:
\begin{equation}
\rho_{ij}=\frac{\langle{\bf R}_{i}{\bf R}_{j}\rangle-\langle{\bf R}_{i}\rangle\langle{\bf R}_{j}\rangle}{\sqrt{\left(\langle{\bf R}_{i}^{2}\rangle-\langle{\bf R}_{i}\rangle^{2}\right)\left(\langle{\bf R}_{j}^{2}\rangle-\langle{\bf R}_{j}\rangle^{2}\right)}}\label{CorrelCoefEq}\end{equation}
where ${\bf R}_{i}$ is the vector of the time series of log-returns.

Each correlation is then converted to a metric distance between pairs of stocks:
\begin{equation}
d_{ij}=\sqrt{2(1-\rho_{ij})}\label{Distance}\end{equation}
forming an $N \times N$ distance matrix $D$. The distance $d_{ij}$ varies between $0$ and $2$, corresponding to correlation values, which can run from $-1$ to $+1$. Small values of $d_{ij}$ thus imply high correlations. The distance matrix is then used to construct the MST.

\section{Results}

We present the results in two sections. We first show the overall static MST, derived from an analysis of the entire sample of data. Following that, a number of dynamic approaches are applied.

\subsection{Static results}

\begin{figure}[H]
\begin{center}
\epsfysize=50mm
\epsffile{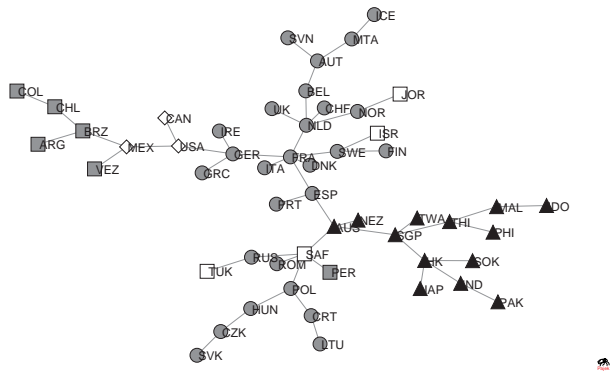}
\caption{Static minimum spanning tree for 1997-2006 for $53$ country equity markets. Coding is: Europe, gray circles (gray $\circ$); North America, white diamonds ($\lozenge$); South America, gray squares (gray $\square$); Asian-Pacific area, black triangles ($\blacktriangle$); and ``other'' (Israel, Jordan, Turkey, South Africa), white squares ($\square$).}
\label{Fig1}
\end{center}
\end{figure}

Shown in Figure \ref{Fig1} is the static MST for the 1997-2006 period. The clusters which we observe appear to be organized principally according to a geographical criterion (possibly also reflecting political and trade criteria). This is similar to the results in \cite{Bonanno_PRE62_7615_2000}. To analyze the graph we identify a ``central'' node, the market most strongly connected to its nearest neighbors in the tree. With the highest number of linkages, France can be considered the central node. Somewhat surprisingly, the US, whose equity market is globally dominant in terms of market value, exhibits a somewhat looser linkage to the other markets. Closely connected to France are a number of the more developed European countries in the European Monetary Union (EMU) and in the EU (Luxembourg, a member of both organizations, is not included in our data set). This European grouping forms a set of markets that are highly correlated with each other, with France at its center. We can also identify several ``branches'' which form the major subsets of the MST and these can then be broken down into ``clusters'' that may or not be completely homogeneous. The Netherlands heads a branch that includes clusters of additional European countries (along with Jordan, anomalously, but which we shall discuss later). The US links a cluster of North and South American countries, except for Peru, to France via Germany. Not surprisingly, the three members of the North American Free Trade Association (NAFTA) - the US, Canada and Mexico - are directly connected, with Mexico forming the link to the South American countries. Australia heads a branch with several groupings: all the Asian-Pacific countries form two clusters, one of more developed and the other of less advanced countries; most of the Central and East European (CEE) countries, that joined the EU in 2004, form an incomplete link to Australia through South Africa, along with Turkey and Peru. Jordan, which appears in a European clustering, is an apparent anomaly. This is likely due to the fact that Jordan is the last node connected to the network and has correlations with other countries close to zero, which means a relatively high minimum distance. We can conclude that Jordan is an outlier of our study that does not have any close relation to any of the other countries represented here.

\subsection{Dynamic results}

The MST presents a static analysis of the relationships between the markets for the time period as a whole. It is possible, however, also to examine the time-dependent properties of the tree to provide insights on the changing relationships between the equity markets over time. To this end several techniques are used. First, we construct  what we call rolling and recursive MST. Second, we show the evolution of the four moments of the mean correlations and  mean tree lengths of the MST. Third, calculation of the mean occupation layer reveals changes in the compactness of the MST over time, the degree of compactness being interpretable as the extent of overall equity market integration. Finally, the single-step and multi-step survival ratios for market linkages provide an indication of the stability of linkages between markets over time.

\subsubsection{Rolling and Recursive MST}

\begin{figure}[H]
\centering
\begin{tabular}{cc}
\begin{minipage}{7.cm}
\includegraphics[width=0.9\textwidth]{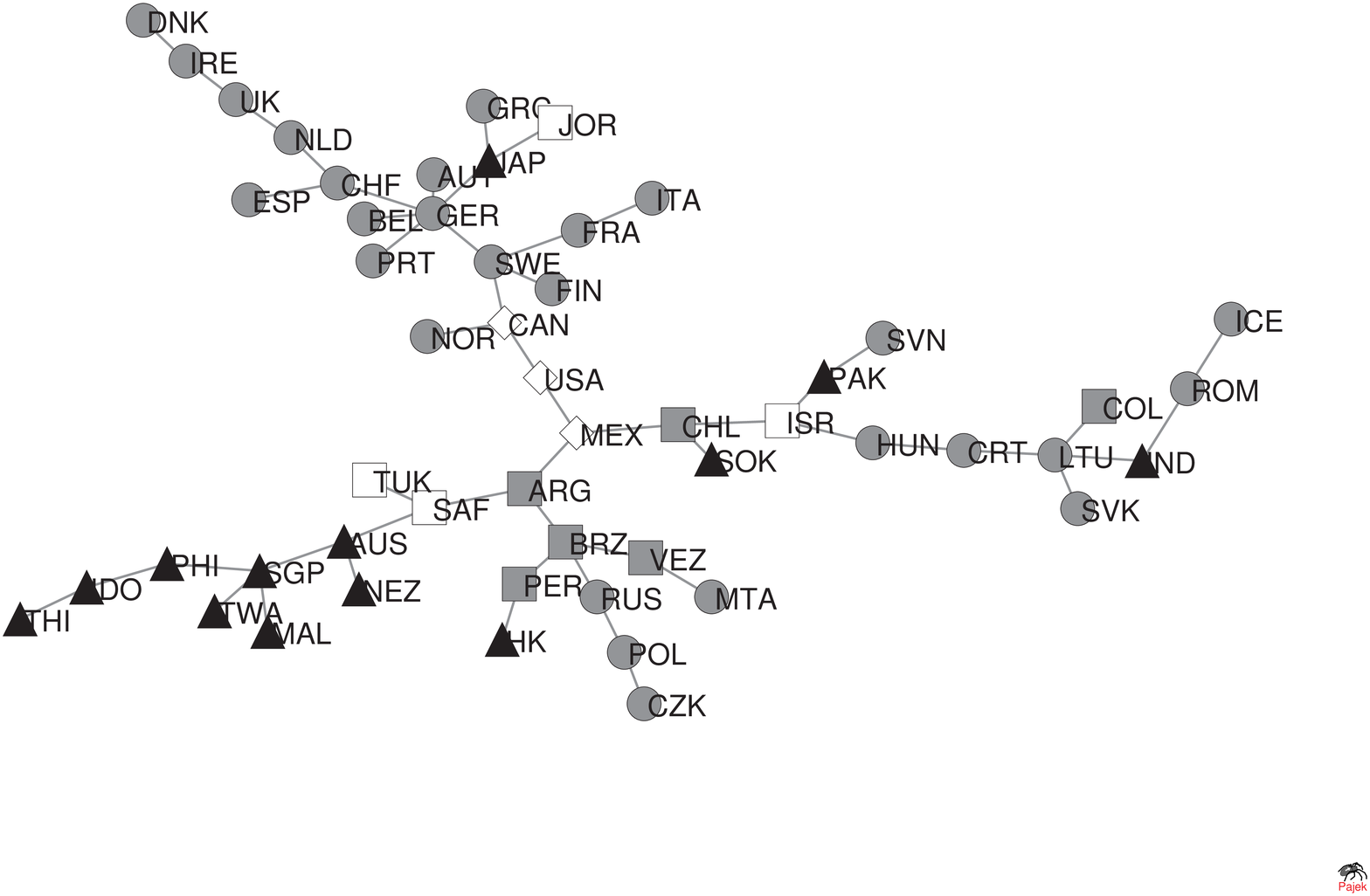}
\begin{center}
1997
\end{center}
\end{minipage}
\begin{minipage}{7.cm}
\includegraphics[width=0.9\textwidth]{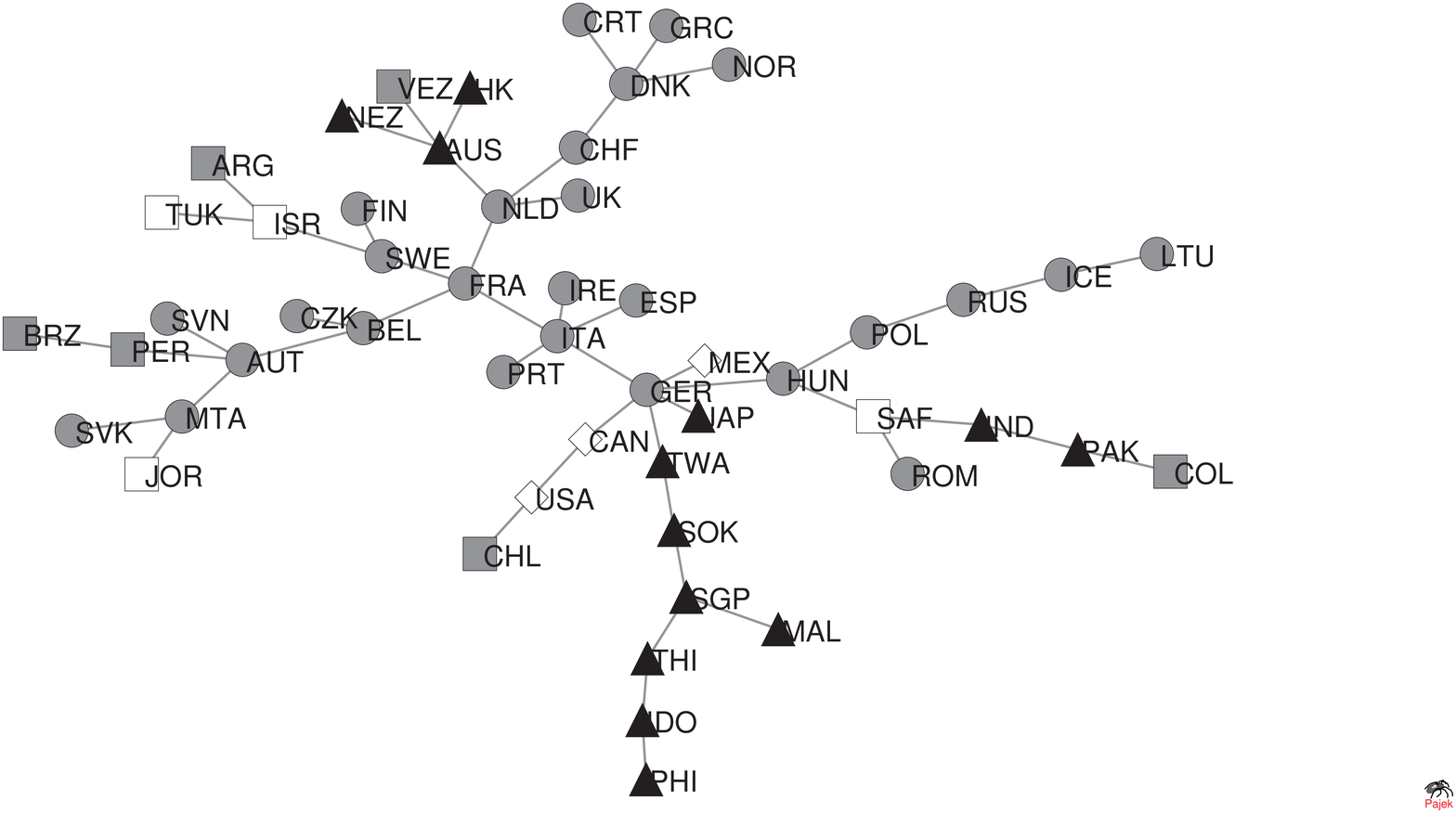}
\begin{center}
2002
\end{center}
\end{minipage}
\end{tabular}
\begin{tabular}{c}
\begin{minipage}{7.cm}
\includegraphics[width=0.9\textwidth]{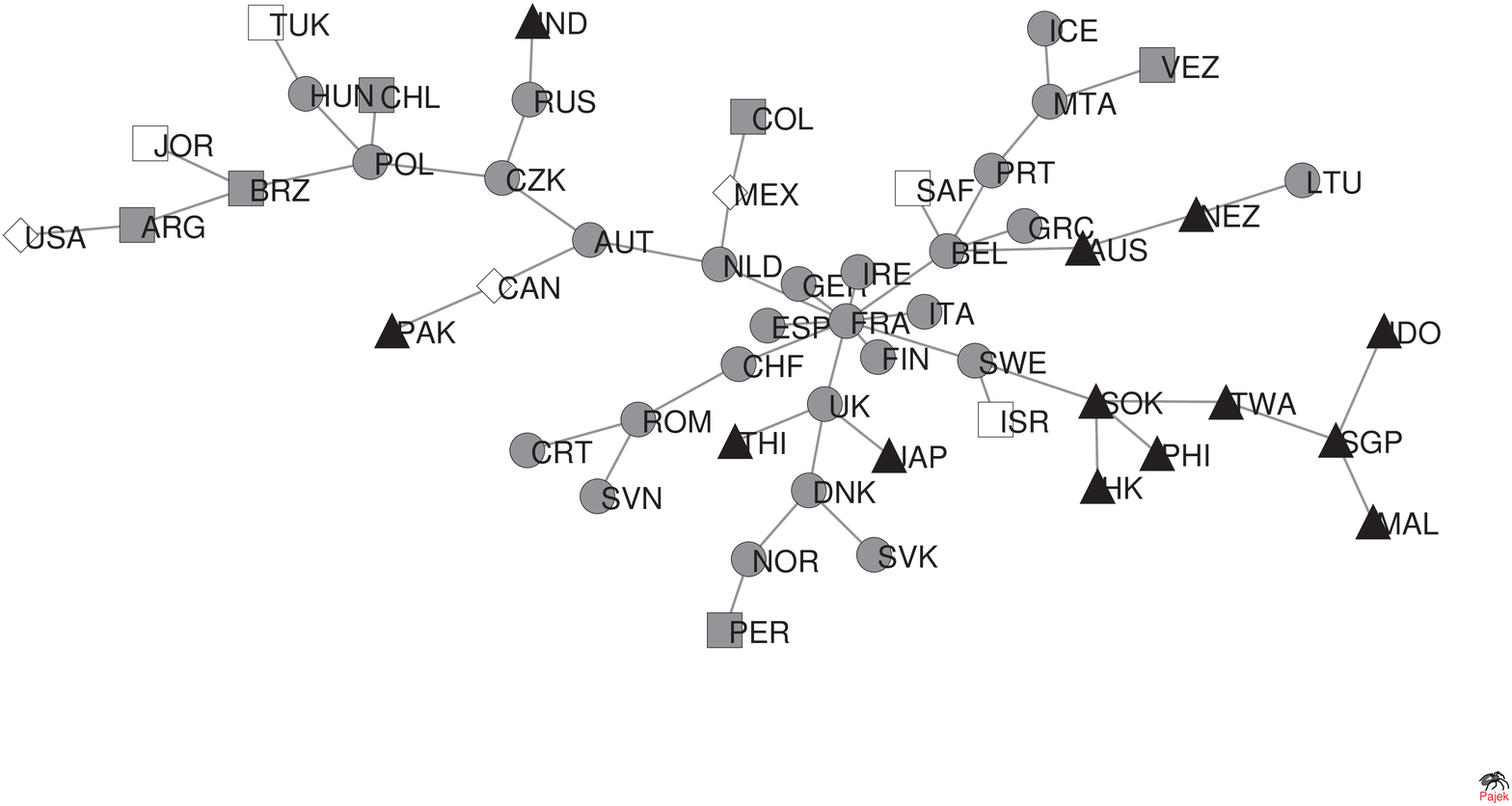}
\begin{center}
2005
\end{center}
\end{minipage}
\end{tabular}
\caption{Rolling one-year window MST for 1997, 2002 and 2005. Coding is: Europe, gray circles (gray $\circ$); North America, white diamonds ($\lozenge$); South America, gray squares (gray $\square$); Asian-Pacific area, black triangles ($\blacktriangle$); and ``other'' (Israel, Jordan, Turkey, South Africa), white squares ($\square$).}
\label{Fig2}
\end{figure}

The dynamic evolution of the MST can be examined by looking at a series of MST created from non-overlapping rolling windows, each with width $T = 1$ year, or $52$ ($53$) weeks. The MST shown in Figure \ref{Fig2} are those for 1997, 2002, and 2005\footnote{The full set of MST graphs is available on request.}.  We detect several consistent relationships as well as a number of less stable, transient, arrangements. One clear consistency is that the developed European countries always form the central structure of the MST. Initially, Germany is the central node; however, in more recent years France has taken over this role. The CEE countries do not form a single cluster but tend to fragment into several subgroups, with changing composition year by year. However, perhaps reflecting the growing economic and political ties with the developed EU members, they tend to move slightly closer to those countries over time in terms of levels away from the central node. With respect to the Asian markets there is usually a link between Australia and New Zealand, which often head a branch connecting most of the remaining Asian markets to Europe. The coherence of the Asian countries is particularly evident in the 1998 MST, possibly reflecting increased correlations in the region in the aftermath of the Asian crisis. This particular clustering does not continue as strongly in subsequent years. The main exception in this group is Japan, which does not fit into the Asian cluster but is generally linked directly to Western markets. Here, trade relationships appear to dominate geographic effects. For the North American markets the US, Canada, and Mexico are usually closely linked, reflecting most likely the ongoing effects of both geography and NAFTA trade ties. An apparent exception is the year 2005. An explanation of the disconnect in this cluster lies in examining the construction of the MST. In 2005 relatively higher correlations between European countries almost completely dominated the formation of the MST as a cluster first formed around France, followed by a group of CEE countries (the Czech Republic, Hungary, and Poland). The South American markets have not formed a complete cluster in any of the years under examination; however, a sub-cluster of Argentina, Brazil, and Chile, the largest, most developed and most liquid Latin American markets, can occasionally be observed. This cluster is usually directly linked to the European grouping via Mexico.

\begin{figure}[H]
\centering
\begin{tabular}{cc}
\begin{minipage}{7.cm}
\includegraphics[width=0.9\textwidth]{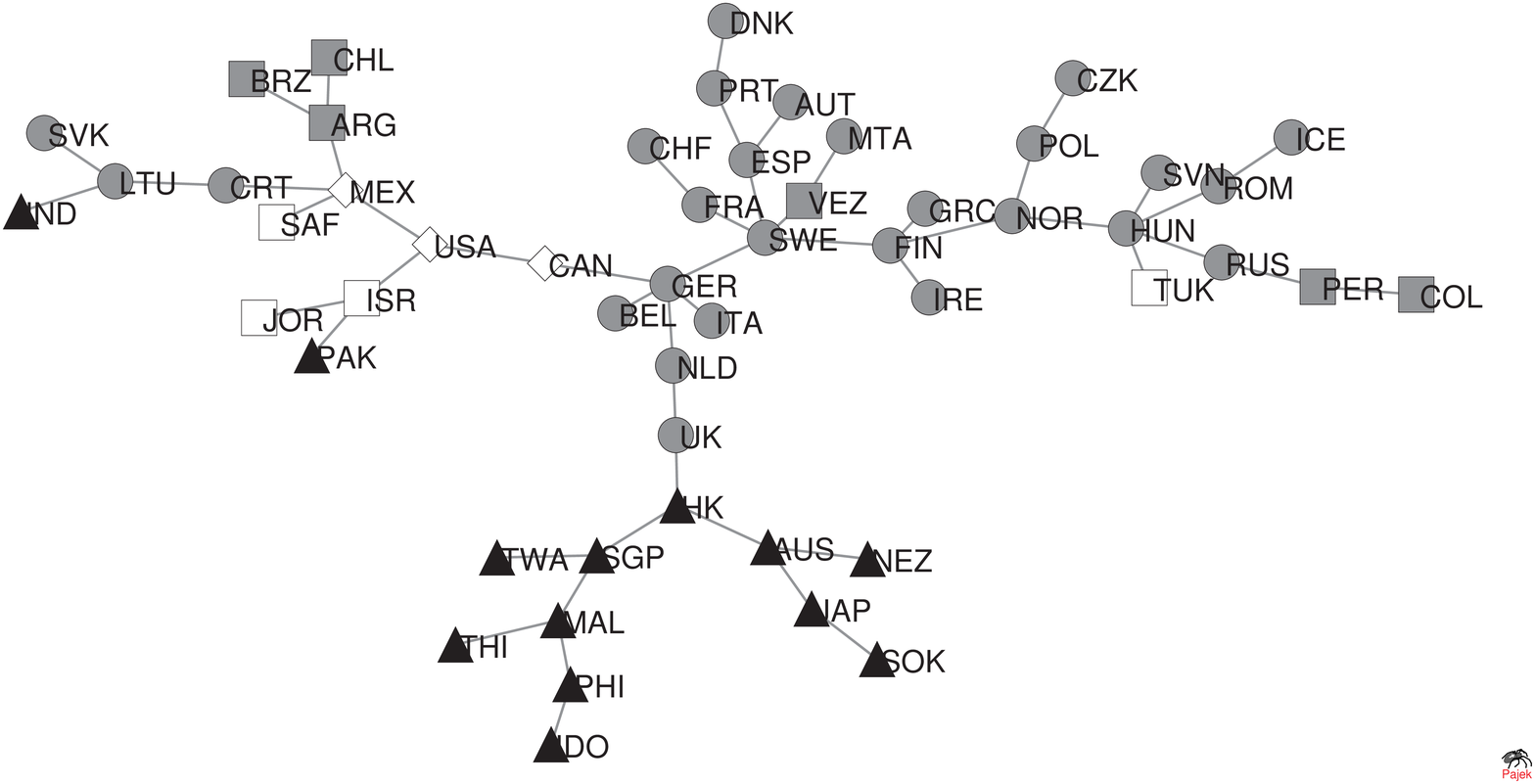}
\begin{center}
1997-1998
\end{center}
\end{minipage}
\begin{minipage}{7.cm}
\includegraphics[width=0.9\textwidth]{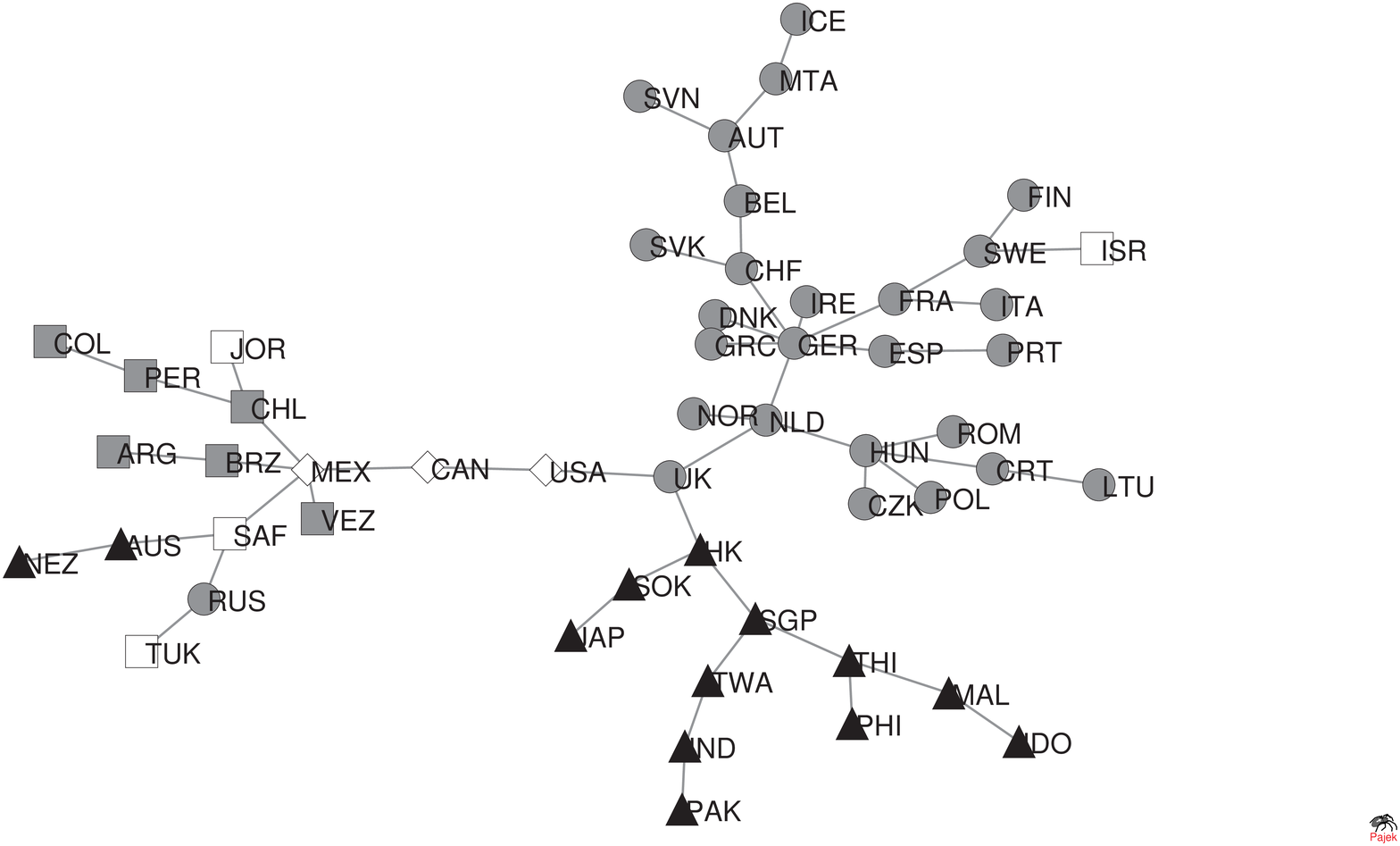}
\begin{center}
1997-2001
\end{center}
\end{minipage}
\end{tabular}
\begin{tabular}{c}
\begin{minipage}{7.cm}
\includegraphics[width=0.9\textwidth]{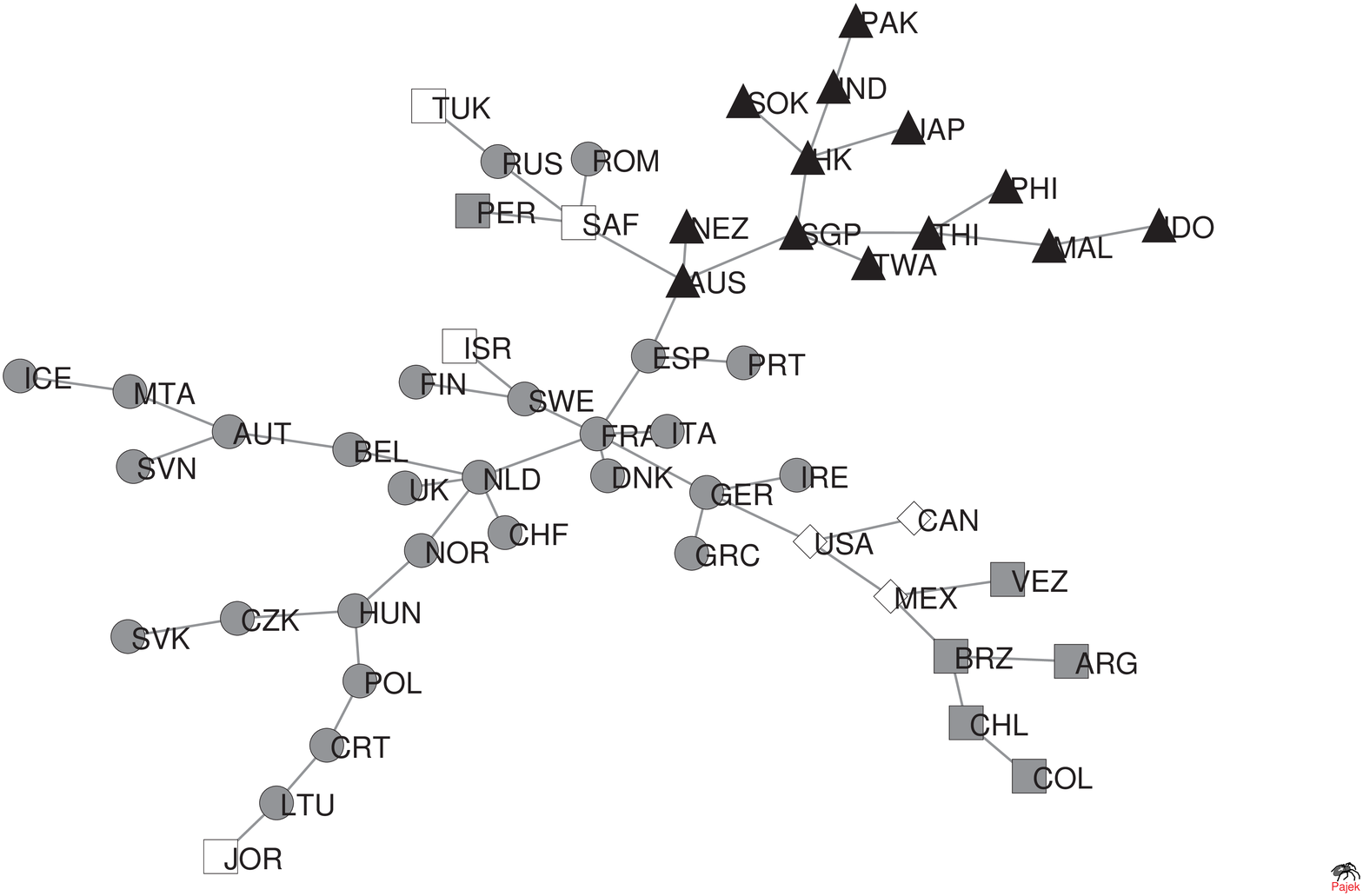}
\begin{center}
1997-2005
\end{center}
\end{minipage}
\end{tabular}
\caption{Recursive MST for cumulative through 1998, cumulative through 2001 and cumulative through 2005. Coding is: Europe, gray circles (gray $\circ$); North America, white diamonds ($\lozenge$); South America, gray squares (gray $\square$); Asian-Pacific area, black triangles ($\blacktriangle$); and ``other'' (Israel, Jordan, Turkey, South Africa), white squares ($\square$).}
\label{Fig3}
\end{figure}

To further examine the stability of the relationships we constructed recursive MST by forming the MST for the first year and then successively adding one year's data at a time. These are shown in Figure \ref{Fig3} cumulatively through 1998, cumulatively through 2001 and cumulatively through 2005 (again, a full set of recursive trees is available on request). Several issues emerge. The first is that the MST appears to have become somewhat more compact in comparison to the rolling window MST for 1997 in Figure \ref{Fig2}. In the 1997 rolling window, the maximum number of levels was twelve (central node Germany to Iceland), while it is consistently smaller in the recursive MST beginning with 1998.  For the period 1997-2005, it is seven (central node France to Jordan). Not surprisingly, given the results from the rolling-window MST, the recursive graphs also reflect the dominance of the developed European grouping and the shift of central node to France from Germany, with Australia, the Netherlands, and the US at the head of the branches. The German-US link persists, even as the center of the European cluster shifts to France. The CEE countries continue to reflect some tendency to split into different clusters, although by 2002 six of them have settled into one group, leaving out only Russia, Slovenia, and Romania. Hungary and Poland, among the more developed CEE equity markets, alternate the role of node linking the CEE countries to the developed EU members. This cluster also gradually moves to a  closer attachment to France as that country becomes the central node. The clustering of the Latin American markets, except for Peru, becomes more consistent as the time period is increased in the recursive graphs, with Mexico generally the link to the European core. Similarly, as the time period increases a more consistent pattern is established for the Asian markets. Pakistan and India join the other Asian countries in 2001, and Australia and New Zealand in 2002. Japan, whose behavior year by year appeared to be largely disconnected from the other Asian markets, is now seen as tied into the Asian cluster via Hong Kong or South Korea consistently since 2001. Finally, a Turkey-Russia-South Africa cluster emerges in 2000 and stays reasonably stable.

\subsubsection{Correlation and mean tree length analysis}

\begin{figure}[H]
\begin{center}
\epsfysize=60mm
\epsffile{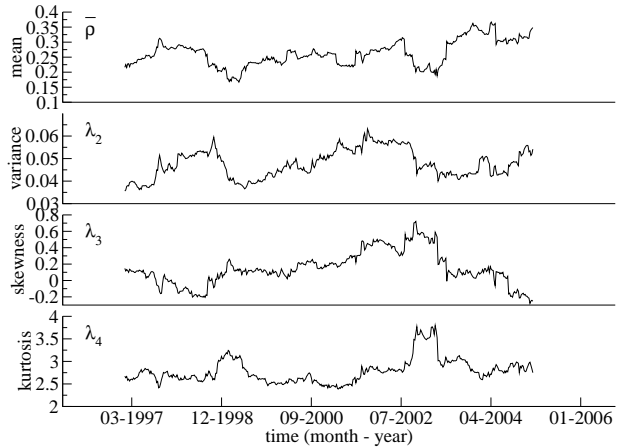}
\caption{Mean, variance, skewness and kurtosis of correlation coefficients as function of time. Window length $T = 52$ weeks and window step length  $\delta T = 1$ week. Results are plotted according to start date of window.}
\label{Fig4}
\end{center}
\end{figure}

\begin{figure}[H]
\begin{center}
\epsfysize=60mm
\epsffile{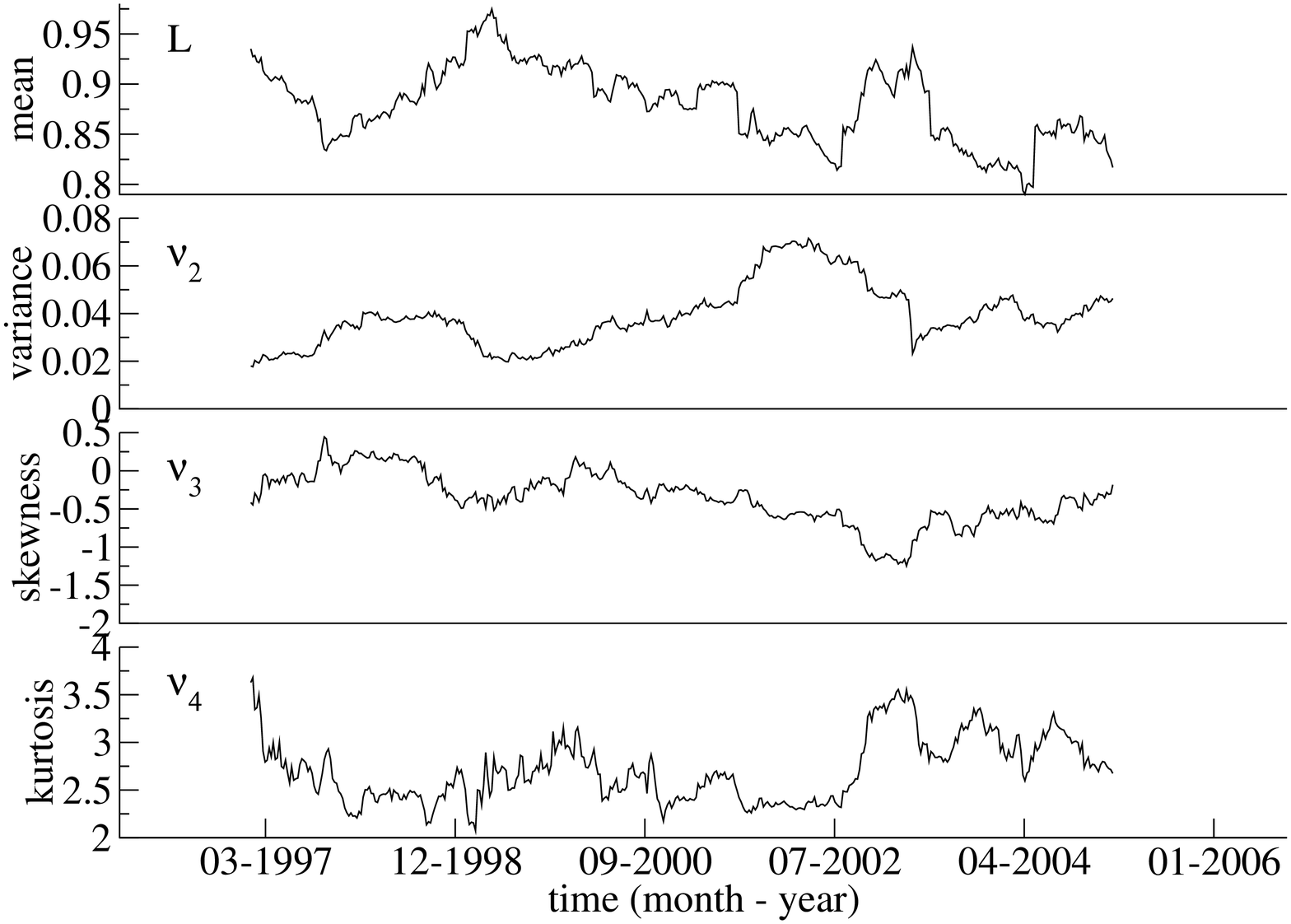}
\caption{Mean, variance, skewness and kurtosis of mean tree length as function of time. Window length $T = 52$ weeks and window step length $\delta T = 1$ week. Results are plotted according to start date of window.}
\label{Fig5}
\end{center}
\end{figure}

Rolling-window graphs of the first four moments of the mean correlations (eq. \ref{CorrelCoefEq}) and of the distances $d_{ij}$ (eq. \ref{Distance}) are presented in Figures \ref{Fig4} and \ref{Fig5}, where the window length is $T = 52$ weeks and the window step length is $\delta T = 1$ week. The mean correlation coefficient is:
\begin{equation}
\overline{\rho}=\frac{2}{N(N-1)}\sum_{i<j}\rho_{ij}.\label{MeanCorrel}\end{equation}
The variance is given as
\begin{equation}
\lambda_{2}=\frac{2}{N(N-1)}\sum_{i<j}(\rho_{ij}-\overline{\rho})^{2},\label{VarianceCorrel}\end{equation}
while the skewness is
\begin{equation}
\lambda_{3}=\frac{2}{N(N-1)\lambda_{2}^{3/2}}\sum_{i<j}(\rho_{ij}-\overline{\rho})^{3},\label{SkewnessCorrel}\end{equation}
and the kurtosis is
\begin{equation}
\lambda_{4}=\frac{2}{N(N-1)\lambda_{2}^{2}}\sum_{i<j}(\rho_{ij}-\overline{\rho})^{4}.\label{KurtosisCorrel}\end{equation}

The moments of the distances $d_{ij}$ in the MST can similarly be calculated over time in terms of the normalized tree length
\begin{equation}
L=\frac{1}{N-1}\sum_{d_{ij}\in\mathrm{\Theta}}d_{ij},\label{MeanLength}\end{equation}
as defined in Onnela {\it et al.} \cite{Onnela_PRE68_056110_2003}, where $N-1$ is the number of edges in the MST. The variance of the normalized tree length is:
\begin{equation}
\nu_{2}=\frac{1}{N-1}\sum_{d_{ij}\in\mathrm{\Theta}}(d_{ij}-L)^{2},\label{VarianceLength}\end{equation}
the skewness is
\begin{equation}
\nu_{3}=\frac{1}{(N-1)\nu_2^{3/2}}\sum_{d_{ij}\in\mathrm{\Theta}}(d_{ij}-L)^{3},\label{SkewnessLength}\end{equation}
and the kurtosis is
\begin{equation}
\nu_{4}=\frac{1}{(N-1)\nu_2^2}\sum_{d_{ij}\in\mathrm{\Theta}}(d_{ij}-L)^{4}.\label{KurtosisLength}\end{equation}

The mean correlation and its variance increase over the initial period covered by the data, roughly corresponding to the era of the Asian and Russian crises. In times of market uncertainty and downturns these measures tend to increase \cite{Onnela_PRE68_056110_2003,Coelho_PH0601189,Baig_IMF_167_1999}. 
The tendency of these measures to increase together has significant implications for standard econometric methodology \cite{Forbes_JF_57_2223_2002}.
We also note that the skewness decreases toward zero, implying that the distribution of the correlations becomes more normal. A similar pattern has been observed for British stocks in the FTSE index \cite{Coelho_PH0601189}. This initial period is followed by declining correlations as global markets move past the 1997-1998 crisis events. Correlations rise again, however, possibly reflecting the broad market declines that begin in 2000. An upward spike occurs in the fall of 2001, which corresponds to the entry into the rolling window of the steepest downturns of global markets as measured by the MSCI world index. Recovery is accompanied, once again, by declining mean correlations. A second, larger, upward spike is observed as the window begins to include the early 2004 period, which occurs in the context of a broader trend toward higher correlations. The mean correlation for weekly returns for the year ending May 13, 2004, is $0.24223$, while that for the year ending May 20, 2004, is $0.30222$. Subsequent correlations remain relatively high. Interestingly, this spike coincides with the entry of new members into the European Union (EU) on May 1, 2004. A breakdown of rolling correlations shows a strong, abrupt increase in correlations for the European group of countries at this point, as well as a consistent tendency over the entire time period for their correlations to be higher than for the set of $53$ markets as a whole. This event has introduced a new element of uncertainty as well as the prospects for closer economic ties, both of which could tend to increase correlations. In contrast to these larger movements the introduction of the euro on January 1, 1999, was not accompanied by major changes in correlation structure.

Essentially the same information provided by the correlation matrix of returns can be obtained also from the graphs of the moments of the distance metric calculated from the MST. The mean distance graph is negatively correlated with the mean correlations, tending to fall, for example, in times of market crisis. This underlines the ability of the MST as a strongly reduced representative of the entire correlation matrix to convey relevant market information. Overall, the mean distance shows a tendency to decrease over the ten years, indicating a ``tighter'' composition of the MST.

\subsubsection{Mean occupation layer}

Changes in the density, or spread, of the MST can be examined through calculation of the mean occupation layer, as defined by Onnela {\it et al.} \cite{Onnela_PRE68_056110_2003}
\begin{equation} 
l(t, v_c) = \frac{1}{N} \sum_{i=1}^N L(v_i^t),
\end{equation}
where $L(v_i^t)$ denotes the level of a node, or vertex, $v_i^t$ in relation to the central node, whose level is defined as zero. The central node can be defined as the node with the highest number of links or as the node with the highest sum of correlations of its links. Using these two definitions, we identify the central node for our rolling-windows with $T = 52$ weeks and $\delta T = 4$ weeks. The two criteria produce similar results. Germany is the central node in the early years, but France takes its place for most of the subsequent periods. Using the highest number of links criterion, France is the central node $41.5\%$ of the time and Germany $27.3\%$. The highest correlation sum criterion identifies France as the central node $53.8\%$ of the time and Germany $30.2\%$. Other countries occasionally assume the position of central node.

\begin{figure}[H]
\begin{center}
\epsfysize=60mm
\epsffile{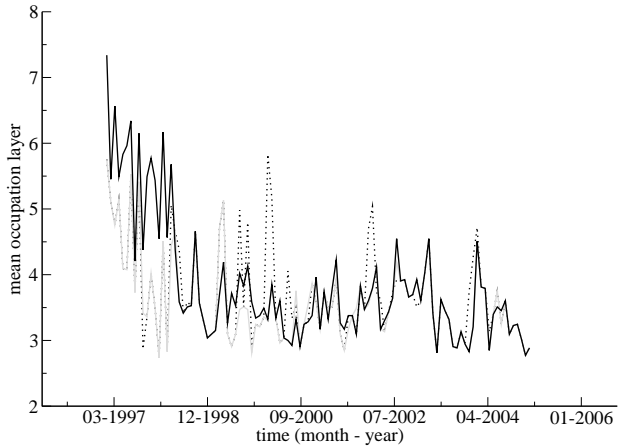}
\caption{Plot of mean occupation layer as function of time ($T = 52$ weeks and window step length $\delta T = 4$ weeks).  Black line shows static central vertex (France), black dots uses dynamic central vertex based on maximum number of links, while gray line shows dynamic central vertex based on maximum correlation value.}
\label{Fig6}
\end{center}
\end{figure}

The mean occupation layer can then be calculated using either a fixed central node for all windows, i.e., France, or with a continuously updated node. In Figure \ref{Fig6} the results are shown for France as the fixed central node (black line), the dynamic maximum vertex degree node (black dots) and the dynamic highest correlation vertex (gray line). The three sets of calculations are roughly consistent. The mean occupation layer fluctuates over time as changes in the MST occur due to market forces. There is, however, a broad downward trend in the mean occupation layer, indicating that the MST over time is becoming more compact.

\subsubsection{Single and Multi Step Survival Rates}

Finally, the robustness of links over time can be examined by calculating survival ratios of links, or edges in successive MST. The single-step survival ratio is the fraction of links found in two consecutive MST in common at times $t$ and $t-1$ and is defined in Onnela {\it et al.} \cite{Onnela_PRE68_056110_2003} as
\begin{equation} 
\sigma (t) = \frac{1}{N-1}|E(t) \cap E(t-1)|
\end{equation}
where $E(t)$ is the set of edges of the MST at time $t$, $\cap$ is the intersection operator, and  $|\cdots|$ gives the number of elements in the set. A multi-step survival ratio can be used to study the longer-term evolution
\begin{equation} 
\sigma(t,k)=\frac{1}{N-1}|E(t) \cap E(t-1) \cdots E(t-k+1) \cap E(t-k)|
\end{equation}
in which only the connections that continue for the entire period without any interruption are counted.

\begin{figure}[H]
\begin{center}
\epsfysize=60mm
\epsffile{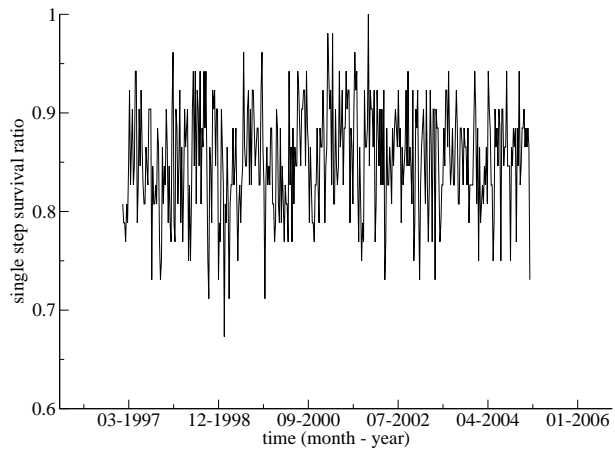}
\caption{Single-step survival ratio as function of time. Window length $T = 52$ weeks and window step length $\delta T = 1$ week.}
\label{Fig7}
\end{center}
\end{figure}

\begin{figure}[H]
\begin{center}
\epsfysize=60mm
\epsffile{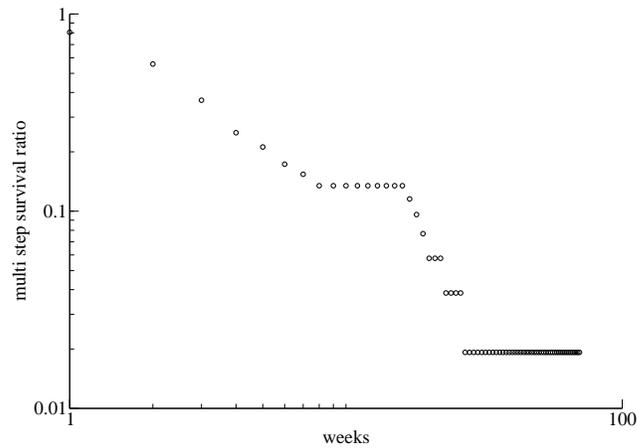}
\caption{Multi-step survival ratio as function of the number of weeks in a log-log scale. Window length $T = 52$ weeks and window step length $\delta T = 1$ week.}
\label{Fig8}
\end{center}
\end{figure}

Figure \ref{Fig7} presents the single-step survival ratios for the MST. The average is about $0.85$, indicating that a large majority of links between markets survives from one window to the next. As might be expected, the ratio increases with increases in window length. Figure \ref{Fig8} shows the multi-step survival ratio. In both cases $T = 52$ weeks and $\delta T = 1$ week. Here, as might be expected, the connections disappear quite rapidly, but a small proportion of links remains intact, creating a stable base for construction of the MST. Again the evidence here is of importance for the construction of portfolios, indicating that while most linkages disappear in the relatively short to medium term there are islands of stability where the dynamics are consistent. The behavior of these two measures is similar to what has been observed for individual stocks within a single equity market \cite{Onnela_PRE68_056110_2003}. These results may understate the stability of the global system of markets since some of the linkage shifts appear to take place within relatively coherent geographical groups. 

\section{Conclusion}

The use of the MST provides a way to extract a manageable amount of information from a large correlation matrix of global stock returns to reveal patterns of links between different markets. It provides an insight into market behavior that is not as easily obtained from the correlation matrix as a whole. Applied dynamically, the analysis lets us observe consistencies as well as evolutions in patterns of market interactions over time. As would be expected, there is a strong tendency for markets to organize by geographical location, although other, related factors such as economic ties, may also play important roles. Developed European countries, with France and Germany at their center, have consistently constituted the most tightly linked markets within the MST. There has also been a limited tendency of the CEE accession countries to link more closely with the more developed EU countries.
 
We have seen that the mean correlations show a tendency to increase over the period as a whole, while mean distances in the MST and the mean occupation layers have been trending downward. These dynamic measures point to a compression of the MST over time, meaning a tighter degree of interaction, or integration, between markets. These findings have implications for the international investor. International diversification under standard Markowitz portfolio construction relies on the existence of a set of assets which display consistent and persistent differences in correlations. These correlations form the basis of the MST. From a Markowitz portfolio perspective \cite{Markowitz_book}, or any portfolio perspective which relies on a spread of (relatively low) correlations, the compression which we have observed implies reduced diversification benefits over the time period we have examined. Finally, the multi-step survival ratio also indicates that while clusters of any given period may be homogeneous, the likelihood of these remaining stable over a reasonable portfolio period is small. This points to a need for frequent restructuring to make maximum use of diversification benefits. 

\begin{ack}
The authors thank the attendees at the IIIS Seminar Series A, esp Jonathan Batten. Gilmore and Lucey acknowledge the support of the Government of Ireland through the Programme for Research in Third Level Institutions. Gilmore also acknowledges support from King's College professional development funds provided under the Senior Faculty Review program. Coelho acknowledge the financial support of Science Foundation Ireland (04/BRG/PO251). This paper was substantially completed while Gilmore was a Visiting Fellow at the IIIS. 
\end{ack}


\begin{thebibliography}{99}

\bibitem{Mantegna_EPJB11_193_1999} R. N. Mantegna, Eur. Phys. J. B \textbf{11}, 193 (1999)

\bibitem{Kearney_IRFA13_571_2004} C. Kearney and B. Lucey, International Review of Financial Analysis \textbf{13}, 571 (2004)

\bibitem{Bekaert} G. Bekaert, C. R. Harvey and R. L. Lumsdaine, Journal of Financial Economics \textbf{65}, 203 (2002) 

\bibitem{Goetzman} W. N. Goetzmann, L. Li and K. G. Rouwenhorst, Journal of Business \textbf{78}, 1 (2005)

\bibitem{Bonanno_QF1_96_2001} G. Bonanno, F. Lillo and R. N. Mantegna, Quantitative Finance \textbf{1}, 96 (2001)

\bibitem{Bonanno_PRE68_046130_2003} G. Bonanno, G. Caldarelli, F. Lillo and R. N. Mantegna, Phys. Rev. E \textbf{68}, 046130 (2003)

\bibitem{Bonanno_EPJB38_363_2004} G. Bonanno, G. Caldarelli, F. Lillo, S. Miccich\`e, N. Vandewalle and R. N. Mantegna, Eur. Phys. J. B \textbf{38}, 363 (2004)

\bibitem{Vandewalle_QF1_372_2001} N. Vandewalle, F. Brisbois and X. Tordois, Quantitative Finance \textbf{1}, 372 (2001)

\bibitem{Onnela_EPJB30_285_2002} J.-P. Onnela, A. Chakraborti, K. Kaski and J. Kert\'esz, Eur. Phys. J. B \textbf{30}, 285 (2002)

\bibitem{Onnela_PhysA324_247_2003} J.-P. Onnela, A. Chakraborti, K. Kaski and J. Kert\'esz, Physica A \textbf{324}, 247 (2003)

\bibitem{Onnela_PST48_106_2003} J.-P. Onnela, A. Chakraborti, K. Kaski, J. Kert\'esz and A. Kanto, Physica Scripta T\textbf{106}, 48 (2003)

\bibitem{Onnela_PRE68_056110_2003} J.-P. Onnela, A. Chakraborti, K. Kaski, J. Kert\'esz and A. Kanto, Phys. Rev. E \textbf{68}, 056110 (2003)

\bibitem{Micciche_PhysA324_66_2003} S. Miccich\`e, G. Bonanno, F. Lillo and R. N. Mantegna, Physica A \textbf{324}, 66 (2003)

\bibitem{Coelho_PH0601189} R. Coelho, S. Hutzler, P. Repetowicz and P. Richmond, \textit{preprint physics/0601189} (accepted to publish in Physica A, article in press)

\bibitem{McDonald_PRE72_046106_2005} M. McDonald, O. Suleman, S. Williams, S. Howison and N. F. Johnson, Phys. Rev. E \textbf{72}, 046106 (2005)

\bibitem{Bonanno_PRE62_7615_2000} G. Bonanno, N. Vandewalle and R. N. Mantegna, Phys. Rev. E \textbf{62}, 7615 (2000)

\bibitem{Panton_JFQA11_415_1976} D. B. Pantom, V. P. Lessig and O. M. Joy, The Journal of Financial and Quantitative Analysis \textbf{11}, 415 (1976)

\bibitem{Schotman_WP04_017_2004} P. Schotman and A. Zalewska, \textit{LIFE Working Paper}, \textbf{04-017} (2004)

\bibitem{Baig_IMF_167_1999} T. Baig and I. Goldfajn, \textit{IMF Working Paper}, \textbf{98-155} (1998)

\bibitem{Forbes_JF_57_2223_2002} K. J. Forbes and R. Rigobon, The Journal of Finance \textbf{57}, 2223 (2002)

\bibitem{Markowitz_book} H. Markowitz, \textit{Portfolio Selection: Efficient Diversification of Investment}, J. Wiley, New York (1959)

\end{thebibliography}
\end{document}